\documentclass[aps,prb,twocolumn,superscriptaddress]{revtex4-2}

\usepackage{siunitx}
\usepackage{graphicx}
\usepackage[caption=false]{subfig}
\usepackage{amssymb}
\usepackage{amsfonts}
\usepackage{amsmath}
\usepackage{amsbsy}
\usepackage[dvipsnames]{xcolor}

\renewcommand{\vec}[1]{\mbox{\boldmath$\mathrm{#1}$}}
\def\ind#1{{_{\mathrm{#1}}}}
\newcommand{\dd}{\mathrm{d}}

\begin{document}
	\title{Topological characterization of dynamic chiral magnetic textures using machine learning}
	\date{\today}
	
	\author{Tim Matthies}
	\affiliation{Institut f\"ur Nanostruktur- und Festk\"orperphysik, Universit\"at Hamburg, 20355 Hamburg, Germany}
	\author{Alexander F. Sch\"affer}
	\affiliation{Institut f\"ur Nanostruktur- und Festk\"orperphysik, Universit\"at Hamburg, 20355 Hamburg, Germany}
	\affiliation{Institut f\"ur Physik, Martin-Luther-Universit\"at Halle-Wittenberg, 06099 Halle (Saale), Germany}
	\affiliation{I. Institut f\"ur Theoretische Physik, Universit\"at Hamburg,	20355 Hamburg, Germany}
	\author{Thore Posske}
	\affiliation{I. Institut f\"ur Theoretische Physik, Universit\"at Hamburg,	20355 Hamburg, Germany}
	\affiliation{The Hamburg Centre for Ultrafast Imaging, 22761 Hamburg, Germany}
	\author{Roland Wiesendanger} 
	\affiliation{Institut f\"ur Nanostruktur- und Festk\"orperphysik, Universit\"at Hamburg, 20355 Hamburg, Germany}
	\author{Elena Y. Vedmedenko} \email{vedmeden@physnet.uni-hamburg.de}
	\affiliation{Institut f\"ur Nanostruktur- und Festk\"orperphysik, Universit\"at Hamburg, 20355 Hamburg, Germany}
	%\author[1]{author4 }
	
	\begin{abstract}
		Recently proposed spintronic devices use magnetic skyrmions as bits of information. The reliable detection of those chiral magnetic objects is an indispensable requirement. Yet, the high mobility of magnetic skyrmions leads to their stochastic motion at finite temperatures, which hinders the precise measurement of the topological numbers. 
		Here, we demonstrate the successful training of artificial neural networks to reconstruct the skyrmion number in confined geometries from time-integrated, dimensionally reduced data. %
		Our results prove the possibility to recover the topological charge from a time-averaged measurement and hence smeared dynamic skyrmion ensemble, which is of immediate relevance to the interpretation of experimental results, skyrmion-based computing, and memory concepts.
	\end{abstract}
	\maketitle

\section{Introduction}
	Large lifetimes of chiral magnetic structures like skyrmions, antiskyrmions, spin helices, or chiral domain walls resulted in numerous experimental and theoretical studies of their application as bits of information in standard \cite{parkin2008magnetic,tomasello2014strategy,hag:NC2015,schaffer2020rotating} and probabilistic computing~\cite{pinna2018skyrmion,zazvorka2019thermal}. The vast majority of these publications relies on the stability of the chiral magnetic objects. Recently, however, it has been shown theoretically \cite{schaffer2019stochastic} and experimentally \cite{song2021commensurability,lindner2020temperature} that skyrmions in various confined geometries can be structurally stable while performing a diffusional motion. Characteristics of this Brownian-like motion are of high importance for memory applications that use confined geometries such as race-tracks or as reservoirs. The dynamics and experimental appearance of skyrmions in this regime of stochastic motion depends on the geometry of the confined regions or islands, temperature, and energy parameters. Importantly, the measured spatial distribution of skyrmions depends on the time resolution of observation. Taking an example of three skyrmions  on a triangular nanoisland, one sees a chaotic skyrmion motion in real-time videos, but three immobile skyrmions in time-averaged static images \cite{schaffer2019stochastic}. Even more peculiar, one sees three immobile skyrmionic features in the time-averaged signal if only two skyrmions are diffusing on a triangular island, but stripes and skyrmion if the island has a round shape. This phenomenon appears particularly important in view of the fact that most experimental techniques resolving chiral magnetic structures are time-integrating. Thus, to rely on chiral magnetic bits of information, it is necessary to distinguish between the spatially and structurally stable magnetic entities from their structurally stable but diffusing counterparts.
	This distinction is, however, not an easy task. Indeed, the required classification of structures cannot be reliably done with the unaided eye, and the direct way to solve it is to increase the time resolution, which is a technically tremendous assignment. Meanwhile, many complex problems have been solved using machine learning techniques. Particularly, machine learning algorithms have been used for finding \cite{Kwon:PRB2019}, recognizing \cite{Iak:PRB2018}, and classifying \cite{Salcedo:JMMM2020} ground states of systems with Dzyaloshinskii-Moriya interactions (DMI) and for reconstructing the spin configuration from data obtained in reciprocal space \cite{Kwon:SciRep2019}. Machine learning has also been used to define phase transitions in Ising-like spin systems \cite{Carras:NatPhys2017} and to estimate the DMI parameters from images of magnetic domains \cite{Kawa:njp2021,kwon2020magnetic}. In contrast to these investigations of statical properties of skyrmionic systems, a recent study \cite{Wang:PRA2021} applied neural networks for deep learning of skyrmionic dynamical phases from videos. Those videos have been created using numerical simulations representing skyrmions as interacting rigid-point particles. Changes in topological charges of chiral structures as well as different time-integrating schemes were disregarded in \cite{Wang:PRA2021}.
	\\
	Here, we show how machine learning categorically outperforms standard advanced techniques for distinguishing diffusing and spatially stable skyrmions and extracting the total topological charge from time-integrated data of the $z$-direction of the magnetic moments. To achieve this goal, we train a neural network (NN) using blurred, time-integrated data in various confined geometries in real space and then calculate topological charges of a complementary set of images using the trained NN. Because the topological charge directly corresponds to the number of topologically non-trivial magnetic objects (skyrmions) in our sample, this procedure delivers complete information about the dynamical system from time-averaged, statical images. Our results yield an outstanding degree of recognition and are important for experimental investigations of stability of chiral magnets.

\section{Skyrmion dynamics in confined geometries}
	We study metastable skyrmions and their motion at finite temperatures on magnetic nanoislands. The simulations are performed with generic parameters close to those of a Pd/Fe bilayer on an Ir(111) surface, known for hosting nanometer-sized skyrmions at moderate magnetic bias fields. A detailed description of the phase diagram for extended systems of ultrathin biatomic layers of Pd/Fe on an Ir(111) substrate hosting magnetic skyrmions can be found in Refs.~\cite{rozsa2016complex,schaffer2019stochastic}. 
	Here, simulations are performed for the same system in the atomistic spin dynamics framework, by solving the Landau-Lifshitz-Gilbert equation (LLG)\cite{gilbert2004phenomenological}
	\begin{align}
		\dot{\vec{m}}_i=&-\frac{\gamma}{1+\alpha^2}\vec{m}_i\times\left(\vec{B}_{i}^{\mathrm{eff}}+\alpha\vec{m}_i\times\vec{B}_{i}^{\mathrm{eff}}\right)\ ,
	\end{align}
	on a discrete square lattice. The gyromagnetic ratio of an electron is  $\gamma=1.76\times 10^{11}($T$^{-1}$s$^{-1})$, $\alpha$ is the Gilbert damping parameter, $\vec{m}_i$ is the normalized magnetization vector at lattice site $i$ and $\vec{B}_{i}^{\mathrm{eff}}$ is the effective magnetic field.
	The classical Heisenberg Hamiltonian reads
	\begin{align}
		\begin{split}
			\mathcal{H} &= -\frac{1}{2}\sum_{\langle ij \rangle}\left[ J\ind{ex} \vec{m}_i\cdot \vec{m}_j -\vec{D}_{ij}\cdot(\vec{m}_i\times \vec{m}_j)\right]\\
			& - \sum_i \left[\mu\ind{s} \vec{B}\ind{ext}\cdot\vec{m}_i - K\ind{u}(m_i^z)^2\right] \, ,
		\end{split}
	\end{align}
	and includes the Heisenberg exchange, interfacial Dzyaloshinskii-Moriya (DMI) interaction, Zeeman coupling to an external field $\vec{B}\ind{ext}$, and the uniaxial anisotropy in $z$-direction. The material parameters (magnetic moment $\mu_\ind{s}=1.037\times10^{-4}~$eV\,T$^{-1}$, interfacial DMI constant $D=1.52~$meV, exchange constant $J\ind{ex}=5.72~$meV, uniaxial anisotropy constant $K\ind{u}=0.4~$meV and Gilbert damping parameter $\alpha = 0.1$) are adopted from Ref.~\cite{Rozsa:PRB2018} while the strength of the  magnetic moment is chosen to be twice lower than that in \cite{Rozsa:PRB2018}. 
	The effective magnetic field is calculated as the functional derivative of the Hamiltonian with respect to the normalized magnetization 
	\begin{align}
		\vec{B}_{i}^{\mathrm{eff}}= -\frac{1}{\mu_s}\frac{\delta \mathcal{H}}{\delta\vec{m}_i} + \vec{B}_{i}^{\mathrm{th}}\, ,
	\end{align}
	in addition to the stochastic thermal field contribution $\vec{B}_{i}^{\mathrm{th}}=\vec{\eta}_i(t)(2\alpha k\ind{B} T / \mu\ind{s}\gamma\Delta t)^{1/2}$. The randomly oriented unit vector $\vec{\eta}$ is reevaluated after each simulation step $\Delta t = 8 \times 10^{-15}\,$s and involves the thermal energy $k\ind{B} T$ (Boltzmann's constant $k\ind{B}$, temperature $T$), hence emulating a stochastic thermal field. If not denoted otherwise, the external parameters are set to $\vec{B}\ind{ext}=1.5\,$T\,$\vec{e}_z$, $T=15\,$K, for which metastable isolated skyrmions are expected to form and move diffusively \cite{schaffer2019stochastic,lindner2020temperature}. 
	
\begin{figure}[htb]
		\centering
		\includegraphics[width=1.0\linewidth]{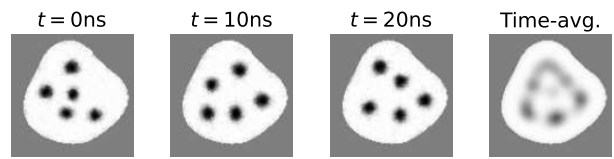}
		\caption{Snapshots of the simulated system at different times and the resulting time-averaged image: In three snapshots the number of skyrmions can be identified to be 5 at all times. Determining this number in the time-averaged image is more difficult.}
		\label{fig:simImagTimeEvo}
	\end{figure}

\begin{figure}[htb]
		\centering
		\includegraphics[width=1.0\linewidth]{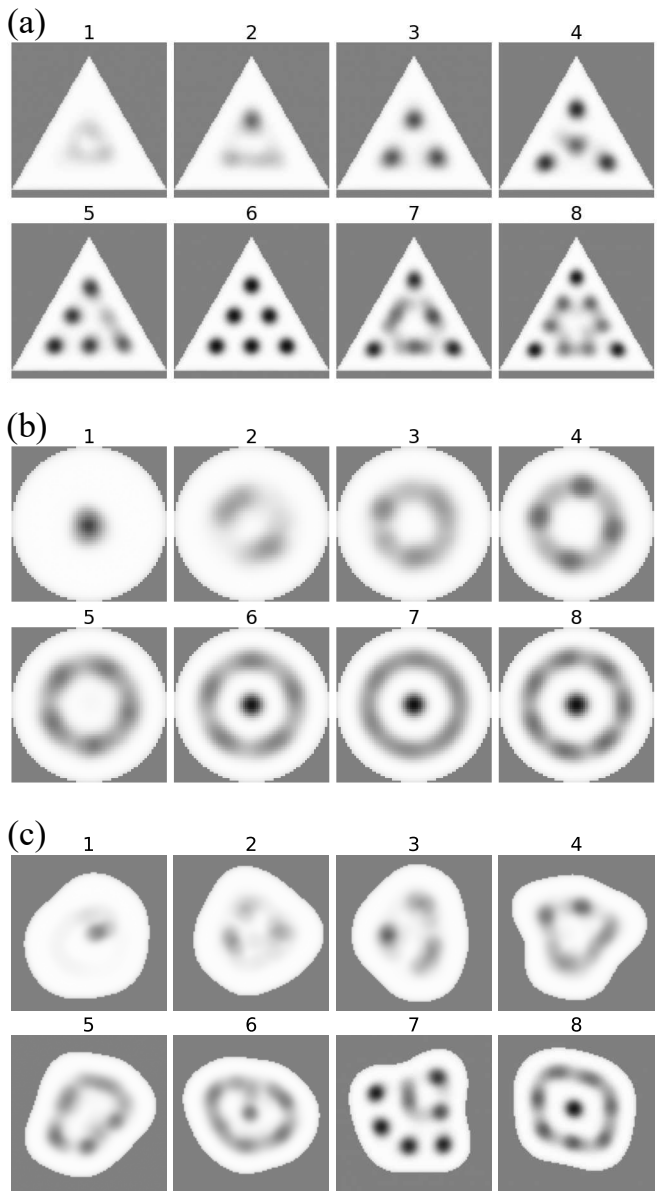}
		\caption{Representative examples of used simulated images with metastable skyrmions for triangular (a), circular (b) and irregular islands (c). The blurred contrast shows the difficulty of predicting the number of skyrmions in a given image. The number of skyrmions is given above the respective figure. The islands are simulated in a rectangle of the size $100\times 100$ atomic sites and the skyrmion size is roughly $15$ atomic sites. Notably, the skyrmion number is assigned correctly by the NN, while counting by hand mostly results in a wrong estimate.}
		\label{fig:simImagsBeforeNorm}
	\end{figure}

	In Ref.~\cite{schaffer2019stochastic} we estimated the time for which a pattern formation in time-integrated measurement techniques stabilizes to be on the scale of $10\,$ns. Hence our simulations are performed over $20\,$ns and the $z-$component of the space-dependent magnetization is averaged over $2.5\times 10^4$ snapshots (cf. Fig.~\ref{fig:simImagTimeEvo}).
	Exemplary resulting patterns with different total topological charges can be seen in Fig. \ref{fig:simImagsBeforeNorm}. As expected, the blurred magnetic contrast prohibits an immediate identification of the skyrmion number with the unaided eye, although the true topological charge can be calculated from the spin structure as 
	\begin{align}
		\label{eq:SkyrmionNumber}
		Q=\frac{1}{4\pi}\iint \vec{m}(x,y)\cdot\left[\partial_x \vec{m}(x,y)\times\partial_y\vec{m}(x,y) \right] \dd x\dd y\, ,
	\end{align}
	for each time-step. This condition is ideal for training an artificial neural network in a supervised learning method. The goal is to overcome the limitations in interpreting time-integrated experimental or simulation data. In the next step, we will elaborate on the design, training and evaluation of the neural network. 
	
\section{Neural network-driven characterization of skyrmion ensembles}
\subsection{Neural Network}

\begin{figure*}
	\centering
	\includegraphics[width=1.0\linewidth]{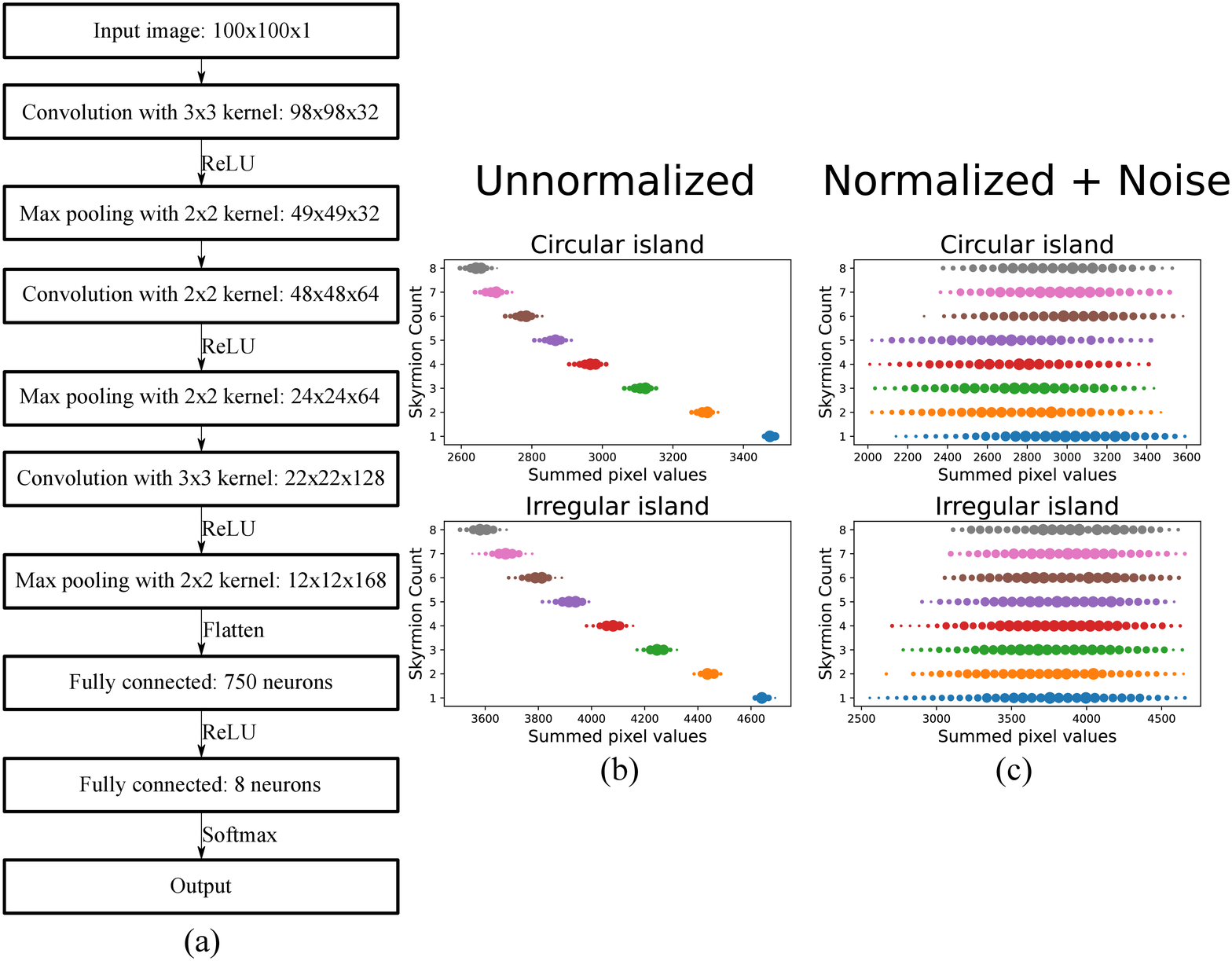}
	\caption{(a) Flowchart showing the network architecture for the used convoluted neural network and scatterplots of the sum over all pixels in a given image: The data set is divided with respect to the number of skyrmions for each configuration. (b) shows the raw data. (c) shows plots where the images are normalized, randomly rescaled and a random noise is added. The size of an individual scatter point scales linearly with the number samples, which exist in its surroundings. Each dot corresponds to an interval with a length of 25 values for (b) and 50 values for (c).}
	\label{fig:NNArchAndScatterAfterNorm}
\end{figure*}

The schematics of our machine learning algorithm is depicted in Fig.~\ref{fig:NNArchAndScatterAfterNorm}(a). The target is the topological number $Q$ corresponding to the vertical magnetization $m_z(x,y)$. In the approach of \cite{BERG1981412}, $Q$ is calculated using all components of the simulated atomic spins. 
However, this calculation can not be performed here by the NN, because only the $z$-components of the spins are known from the experimental measurements. Here, the NN hence has to predict $Q$ solely on the information given by the time-averaged $z$-projection of the spins $\bar{m}_z(x,y)=\langle m_z(x,y)\rangle_t$.
The $\bar{m}_z(x,y)$ map is used as the input layer. The specifics of the used data augmentation and preprocessing on the input will be explaind later.
Interestingly, this reduced information is insufficient to  reconstruct the skyrmion number by mathematical means, i.e., without knowledge about the in-plane rotational sense of the magnetic moments, the skyrmion number in  Eq.~(\ref{eq:SkyrmionNumber}) can not be calculated per se. It is hence astonishing that the neural network is capable of correctly assigning the topological number from the out-of-plane magnetization alone. The reason for this success lies in the specifics of the system, which exclusively hosts non-overlapping localized skyrmions with a fixed sense of rotation. By identifying these structures, the network is able to effectively reconstruct the missing directional information and assign the skyrmion number correctly.    
The output layer corresponds to a category vector with $8$ entries. Each entry corresponds to a topological number between $1$ and $8$. Since the case $Q=0$ is analogous to an empty island with no magnetic structure after time-integration, the classification is trivial. Therefore, $Q=0$ is omitted from the data set. For the systems examined here, the topological number $Q$ is identical to the number of localized skyrmions in the system, because with the used simulation parameters the skyrmions are well separated.
The maximum value of $Q$ is 8, because when placing more than 8 skyrmions in the island, our simulations show that skyrmions would escape at the edge of the island dynamically. In general, the range of $Q$ can be approximated in advance by dividing the area of the island by the equilibrium skyrmion area and the characteristic spacing between the skyrmions in the system.
The key point is to train the NN to statistically learn the relationship between the time-averaged magnetic contrast along the $z$-axis (Input) and the topological number $Q$ (Output), which is time-independent for the samples evaluated here. The network architecture consists of a convolutional neural network, with three convolutional layers, each followed by a maximum pooling layer and a fully connected layer at the end. The number of filters for the three convolutional layers are 32, 64, and 128,  respectively. After each convolution step and after the fully connected layer the ReLU activation function is applied. The output layer consists of 8 neurons, each corresponding to a different amount of skyrmions in the configuration. Categorical cross-entropy is chosen for the loss function
\begin{equation}
	\mathcal{L}(\mathbf{y},\mathbf{\hat{y}})=- \frac{1}{N} \sum_{i=0}^N \sum_{j=1}^8 y_{ij}\, ln(\hat{y}_{ij})\ ,
\end{equation}
where $N$ denotes the batch size (amount of data over which one training iteration is performed), $\mathbf{y}$ is the category vector of the labels, and $\mathbf{\hat{y}}$ is the prediction. At the end of the NN the softmax function is applied to the output yielding a probability distribution for $\mathbf{\hat{y}}$
\begin{equation}
	\hat{y}_{ij} = \frac{e^{z_{ij}}}{\sum_{k=1}^{8}e^{z_{ik}}}\ .
\end{equation}
$\mathbf{z}$ is the vector corresponding to the last layer. As the final prediction of the network, we choose the Q corresponding to the neuron with the highest output signal. The batch size is set to 32. The Adam optimizer~\cite{kingma2015ICLR} with a learning rate of 0.001, is used to update the weights and biases of the NN. The custom architecture used here is similar to that of LeNet-5~\cite{lecun1998leNet}. A comparable custom architecture has been used for other analyses of magnetic systems~\cite{Kawa:njp2021,Salcedo:JMMM2020}. The network structure was chosen after varying the number of convolutional and fully connected layers, the filter and kernel size, the number of neurons, and the activation function. This architecture resulted in a stable and fast training process.
The NN was implemented in TensorFlow~\cite{tensorflow2015-whitepaper}. The training was performed on Google Colaboratory~\footnote{on Google Colaboratory, available at: https://colab.research.google.com/}.

\subsection{Training}
	The simulations are performed for three different kinds of island geometries: Circular, triangular and irregular islands.  A sample of the generated data is shown in Fig.~\ref{fig:simImagsBeforeNorm}. The generated data is then split into training and test data sets with 40\% of the data used for testing as shown in table ~\ref{tab:trainingTestQuantities}.

	\begin{table}[h]
		\caption{\label{tab:trainingTestQuantities}Size of the data sets for three classes of islands without data augmentation}
		\centering
		\begin{ruledtabular}
			\begin{tabular}{ccc}
				Island & Size of training set & Size of testing set\\ \hline
				Circular & 3622 & 2415\\
				Triangular & 3030 & 2035\\
				Irregular & 3584 & 2404\\
			\end{tabular}
		\end{ruledtabular}
	\end{table}
	Through sample flipping and four-fold rotation, the size of the training data set is increased by a factor of 8. 
	
\begin{figure}[h!]
		\centering
		\includegraphics[width=1.0\linewidth]{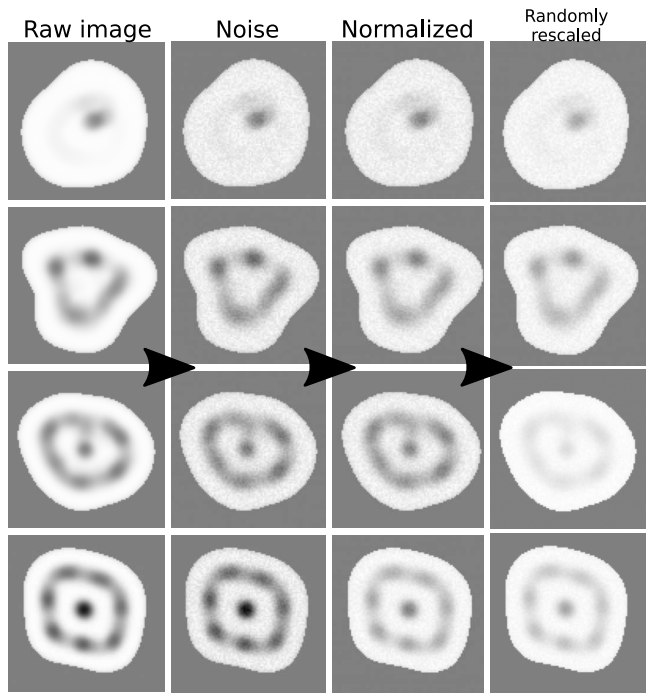}
		\caption{Image processing of the data: The images are normalized, randomly rescaled and injected with noise. This is done to prevent a classification solely on net magnetization.}
		\label{fig:ImgProcessing}
	\end{figure}

\begin{figure}[h!]
		\centering
		\includegraphics[width=1.0\linewidth]{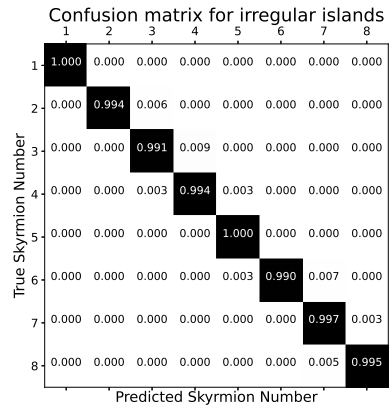}
		\caption{Confusion matrix of the convolutional neural network for irregular islands: This matrix shows the distribution of the predictions of the network. Each row is normalized, giving a probability distribution for the prediction over all samples with the same number of skyrmions. In total $99.5\%$ of the samples are classified correctly. }
		\label{fig:ConfMatIreg}
	\end{figure}

	Since each skyrmion contributes a constant amount of magnetization in the $z$-direction, the simulated data can be classified by only looking at the net magnetization. This is shown in Fig.~\ref{fig:NNArchAndScatterAfterNorm}(b). For each image in the data set, the brightness value of each pixel is summed up. This quantity, which is proportional to the net magnetization, is plotted as a function of the number of skyrmions in the system. The gaps between the data points enable the classification without the NN. To ensure that the data represents experimentally obtained data more closely, different steps of image processing are applied:
	\begin{equation}
		\Bar{x}_{ij}= \frac{x_{ij} + 0.2 \epsilon_{ij}}{\max({x_{ij}+ 0.2 \epsilon_{ij}})}\cdot (0.8 \eta + 0.2)\ .
	\end{equation}
	Here, $\mathbf{x}$ and $\mathbf{\Bar{x}}$ are the pixel values before and after the image processing respectively, $\mathbf{\eta}$ is a uniformly distributed random variable with $\eta \in [0,1]$, and $\mathbf{\epsilon}$ is uniformly distributed noise with $\epsilon_{ij} \in [0,1]$.
	This forces the NN to utilize the patterns which form in the configurations and impedes the classification on net magnetization. The example images after each step are shown in Fig.~\ref{fig:ImgProcessing}.
	After the training process, 100\% of the testing data for the triangular islands are correctly identified. For that, the network is trained for $10$ epochs. One epoch is a training iteration over the whole training data set. For the irregular and the circular islands the network is capable of predicting 99.5\% of the testing data set correctly, as can be seen in the confusion matrix in Fig.~\ref{fig:ConfMatIreg}. The matrix shows the predicted skyrmion number as a function of skyrmions in the system. Due to the regularity of the patterns formed in the triangular and circular islands, conventional machine learning methods such as k-nearest neighbors (k-NN) and support vector machines (SVM) can achieve similar results. For the irregular islands, these methods fall significantly behind our approach based on artificial neural networks. Here, we find a maximal accuracy for k-NN of 32\% and a maximal accuracy for SVM of 61\% when applied to irregularly shaped islands.
	
\section{Discussion}
	In this paper we showed the successful training of an artificial neural network to correctly identify stochastically moving ensembles of skyrmions in confined geometries. The data sets are obtained from classical spin dynamics simulations, mimicking measurement techniques that are sensitive to the out-of-plane component of the magnetization exclusively. Additionally, the time resolution is below the characteristic skyrmion dynamics, which leads to mixtures of blurred out skyrmion traces and localized skyrmions. Though, these patterns do not allow an identification of the true skyrmion number with the unaided eye, nor by analytical means, as this would demand time-resolved data of the magnetization for all three directions. The NN, trained with supervised learning, grants a topological charge identification with vast reliability.
	Future experiments on the dynamics of magnetic configurations can benefit from these findings immediately, which help to identify the magnetic configuration at hand, even for highly mobile quasiparticles whose velocity exceeds the experimentally limited time resolution.
	Our results show a further possibility of benefiting from state-of-the-art machine learning techniques. Possible next steps include the generalization of the pattern recognition to different non-collinear magnetic textures. Especially coexisting localized and delocalized skyrmions and spin spirals are relevant cases that are hard to characterize with conventional methods.
	It should be noted that we train the network for the physical parameters that describes the physical system best. To account for different physical systems, e.g., different materials or strongly altered system size and geometry, the training dataset must be adjusted accordingly. For example, if we doubled the island size and thus decrease the relative skyrmion size, the network trained with the original data fails to predict the correct topological charge Q. The patterns that form in an island twice the size are too different.
	Finally, the correct identification of ensembles of skyrmions at finite temperatures lays the basis for realizing skyrmion-based applications in conventional or unconventional computing applications like reservoir, stochastic or neuromorphic computing. 
    
    \section*{Acknowledgments}
        A.F.S. acknowledges the Universität Hamburg's funding line Next Generation Partnerships funded under the Excellence Strategy of the Federal Government and the Länder, as well as financial support from the German research foundation (DFG) through the collaborative research center CRC/TRR 227. T.P. and R.W. acknowledge funding by the DFG via the Cluster of Excellence ''Advanced Imaging of Matter'' (EXC 2056, Project ID 390715994).

\end{document}